\documentclass[a4paper,11pt]{article}
\pdfoutput=1 

\usepackage{jinstpub} 
\usepackage{lineno}

\usepackage{float}
\usepackage{graphicx}
\graphicspath{{figures/}}
\usepackage{caption}
\usepackage{subfig}

\usepackage{tikz}
\usetikzlibrary{positioning}
\tikzset{>=stealth}
\usepackage{xcolor}
\usepackage{overpic,color}
\usepackage{caption3} 
\DeclareCaptionOption{parskip}[]{} 

\usepackage[mathscr]{eucal}
\definecolor{LightRed}{RGB}{255,102,102}
\definecolor{DarkRed}{RGB}{153,0,0}
\definecolor{LightGreen}{RGB}{153,255,51}
\definecolor{DarkGreen}{RGB}{0,102,0}
\definecolor{LightBlue}{RGB}{99,184,255}
\definecolor{DarkBlue}{RGB}{0,0,102}
\definecolor{DarkYellow}{RGB}{255,171,0}
\definecolor{shadecolor}{rgb}{0.92,0.92,0.92}
\usepackage[normalem]{ulem} 
\newcommand\hl{\bgroup\markoverwith
	{\textcolor{yellow}{\rule[-.5ex]{2pt}{2.5ex}}}\ULon}
\usepackage{hyperref}

\title{\boldmath Scintillator Tile Batch Test of CEPC AHCAL}

\author[b,a]{Y. Duan,}
\author[e,f,i]{J. Jiang,}
\author[a]{J. Li,}
\author[g,h]{L. Li,}
\author[b,a,1] {S. Li,\note{Corresponding author.}}
\author[b,a]{D. Liu,}
\author[c,d]{J. Liu,}
\author[e,f,i]{Y. Liu,}
\author[e,f,i]{B. Qi,}
\author[b,a]{R. Qian,}
\author[c,d]{Z. Shen,}
\author[c,d]{Y. Shi,}
\author[a]{X. Wang,}
\author[b,a]{Z. Wang,}
\author[a,b,1] {H. Yang,}
\author[e,f,i]{B. Yu,}
\author[c,d]{and Y. Zhang}

\affiliation[a]{Institute of Nuclear and Particle Physics, School of Physics and Astronomy, Key Laboratory for Particle Physics and Cosmology, Shanghai Jiao Tong University, \\Shanghai 200240, P.R. China}
\affiliation[b]{Tsung-Dao Lee Institute, Shanghai JiaoTong University, \\Shanghai, 200240, P.R. China}

\affiliation[c]{State Key Laboratory of Particle Detection and Electronics, University of Science and Technology of China, \\Hefei 230026, P.R. China}
\affiliation[d]{Department of Modern physics, University of Science and Technology of China, \\Hefei 230026, P.R. China}
\affiliation[e]{Institute of High Energy Physics, Chinese Academy of Sciences (CAS), \\Beijing 100049, P.R. China}
\affiliation[f]{University of Chinese Academy of Sciences, \\Beijing 100049, P.R. China}

\affiliation[g]{Department of Nuclear Science and Technology, College of Material Science and Engineering, Nanjing University of Aeronautics and Astronautics, \\Nanjing, Jiangsu 211106, P.R. China}
\affiliation[h]{Institute of Nuclear Analytical Technology, Nanjing University of Aeronautics and Astronautics, \\Nanjing, Jiangsu 211106, P.R. China}
\affiliation[i]{State Key Laboratory of Particle Detection and Electronics, Institute of High Energy Physics, \\Beijing 100049, P.R. China}
\emailAdd{yduan@sjtu.edu.cn}

\abstract{Hadron calorimeter (HCAL) is an essential sub-detector of the baseline detector system for Circular Electron Positron Collider (CEPC). We plan to build an Analog Hadron CALorimeter (AHCAL) prototype based on the Particle Flow Algorithm (PFA). The AHCAL of CEPC uses steel as absorber and scintillator tiles read out by Silicon Photo-Multipliers (SiPMs) as sensitive medium. The energy linearity and resolution of the calorimeter depends on the light yield uniformity of sensitive medium. It is essential to qualify the entire detector production in order to select scintillator tiles with light yield uniform within 10\%. An automated batch test platform has been designed with 144 channels, an automated 3D servo motor. The paper summarizes the tests performed on more than 15000 scintillator tiles. The measured light yield, corrected for the set-up response non-uniformity, is around 12.9 p.e. .  About 91.6\% of scintillators (14219 pieces) are qualified within 10\% of light yield window.}

\keywords{CEPC; AHCAL; PFA; SiPM; Scintillator}

\begin{document}
\maketitle
\flushbottom

\section{Introduction}
\label{sec:intro}
The discovery in July 2012 by the A Toroidal Large Hadron Collider ApparatuS (ATLAS) ~\cite{a} and Compact Muon Solenoid (CMS) experiments ~\cite{b} at the Large Hadron Collider (LHC) ~\cite{c} of a new particle with mass of 125 GeV and properties compatible with the Higgs boson ~\cite{d,e} predicted by the Standard Model (SM) ~\cite{f} is a major milestone in High Energy Physics. Chinese Higgs factory community proposed the Circular Electron Positron Collider (CEPC) ~\cite{g,h,i}, serving also as W and Z factory. \par
CEPC serves as Higgs, W and Z factories, it is desired to achieve an unprecedented jet energy resolution of 30\%/$\sqrt{E(GeV)}$. The CEPC detector is composed of a precision vertex detector, a time projection chamber, a silicon tracker, a muon detector and a high granularity calorimetry system. The calorimetry subsystem consists of an electromagnetic calorimeter (ECAL) and a hadronic calorimeter (HCAL). The traditional calorimetry can hardly achieve the energy resolution requirement, the high granularity calorimetry system based on the particle flow algorithm (PFA ~\cite{j,k}) is selected as a baseline detector, which is highly segmented to separate close-by shower particles by an excellent 3D positioning resolution and reconstruct the individual particle showers. The main idea of PFA is the combination of the jets' final state measurement, which is 65\% charged hadron measured by tracking system, 25\% photon by ECAL and 10\% neutron hadron by ECAL+HCAL. The high granularity calorimetry system is required to provide a energy resolution for roughly 16\%/$\sqrt{E(GeV)}$ for the photon and 60\%/$\sqrt{E(GeV)}$ for the neutral hadron. \par
\begin{figure}[H]
	\centering 
	\includegraphics[width=.9\textwidth]{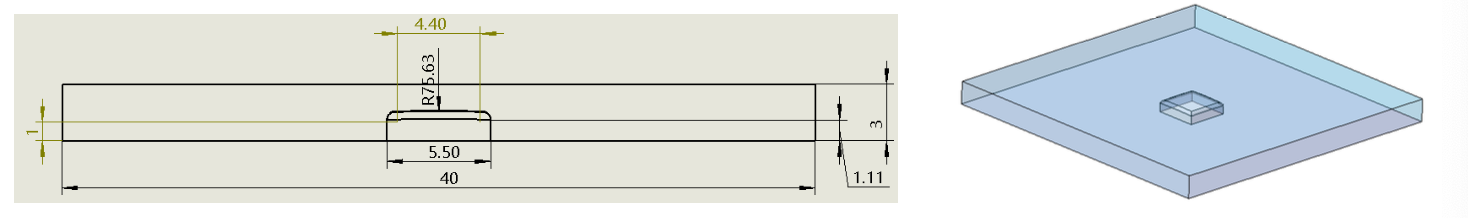}
	\caption{\label{fig:ii} SiPM-on-Tile design. The reported thicknesses are ex- pressed in mm.}
\end{figure}%
HCAL R\&D project plan to build a 40 layers prototype, of dimensions 70 cm $\times$ 70 cm $\times$ 100 cm, using steel as absorber and plastic scintillator ~\cite{sci} read out by a SiPM ~\cite{sipm} as sensitive medium. The SiPM-on-Tile is shown in figure \ref{fig:ii} (for more information see ~\cite{sci}). The scintillator with a dimension 40mm$\times$40mm$\times$3mm are made of polystyrene by the injection molding technique and wrapped in ESR films. Light yield is defiled as the capability of scintillators to produce photons from deposited energy. It depends on production and wrapping of the scintillators. SiPMs can turn the photons to photoelectrons and magnify them, then the multiplied photoelectrons are collected and amplified by SPIROC2E ~\cite{SP14}, a 36-channel front-end chip which could sample and digitize the input signals.\par
The non-uniformity among different channels will eventually deteriorate the linearity and resolution of the energy reconstruction ~\cite{l}. The energy resolution of the reconstructed particles is proportional to the non-uniformity. There are three aspects that can affect the non-uniformity: (1) scintillator light yield, (2) SiPM gain, (3) SPIROC2E electronic gain. This article will report about tests performed on more than 15000 scintillator tiles produced for the realization of the AHCAL prototype.
\section{Batch test platform}
\label{sec:BatchT}
\begin{figure}[H]
	\centering
	\subfloat[]{\label{fig:iii}\includegraphics[width=.6\textwidth]{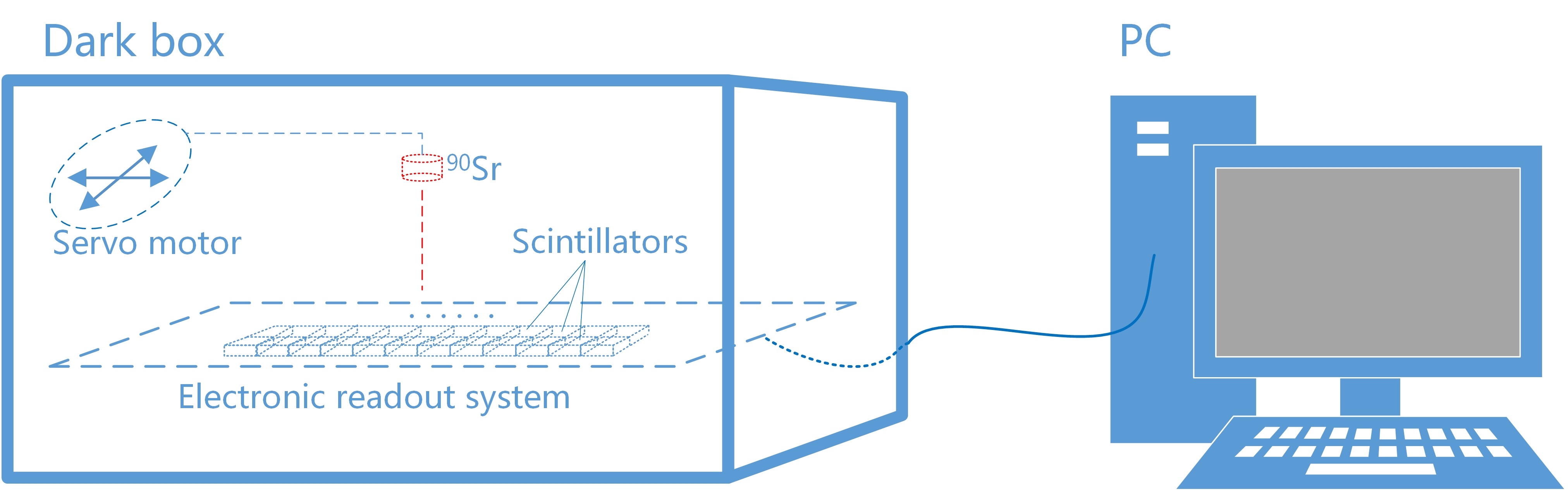}}
	\subfloat[]{\label{fig:i}\includegraphics[width=.4\textwidth]{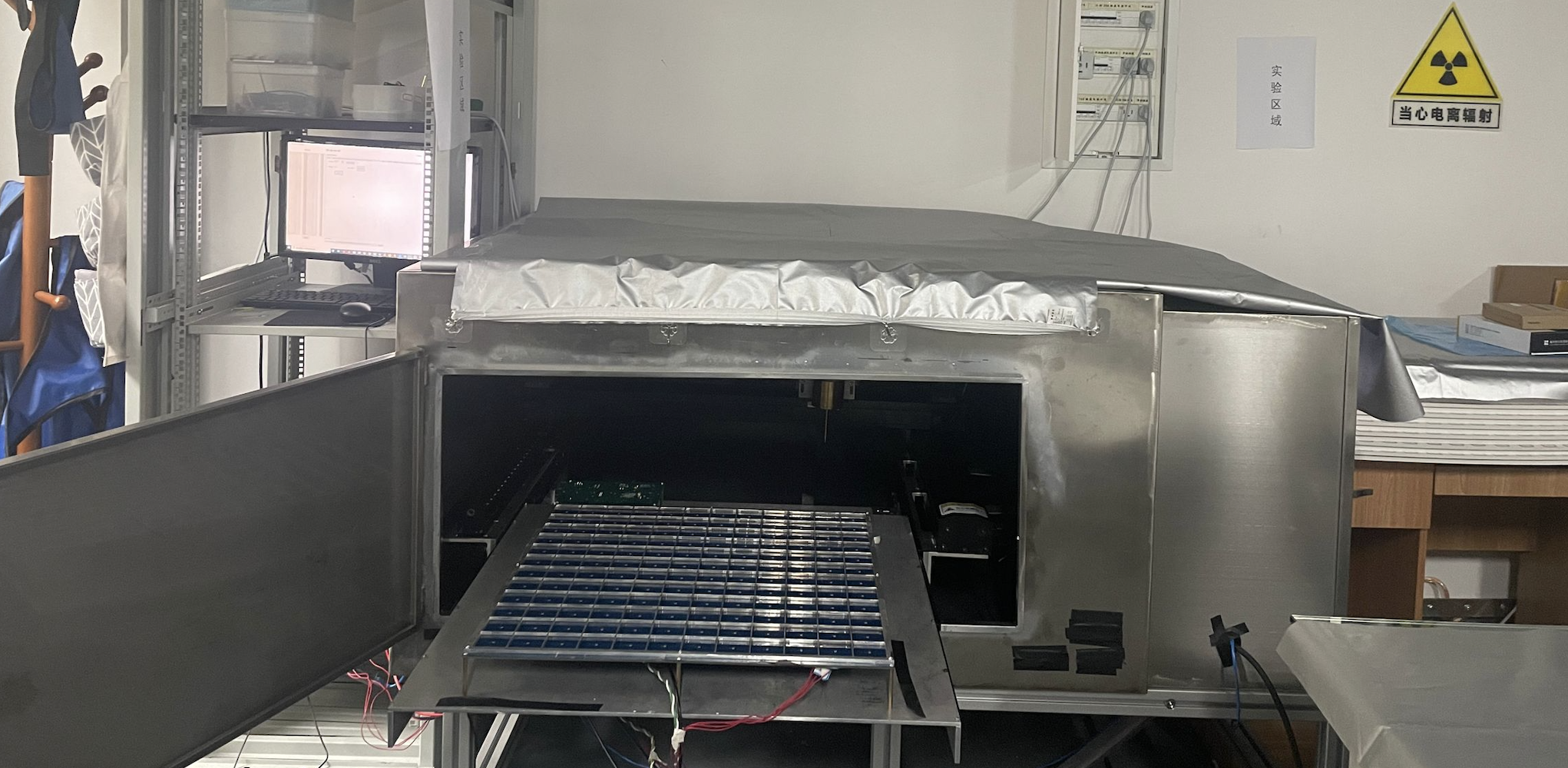}}
	\caption{\label{fig:iiii} Batch test platform sketch (a) and picture (b).}
\end{figure}
\begin{figure}[H]
	\centering
	\includegraphics[width=.6\textwidth]{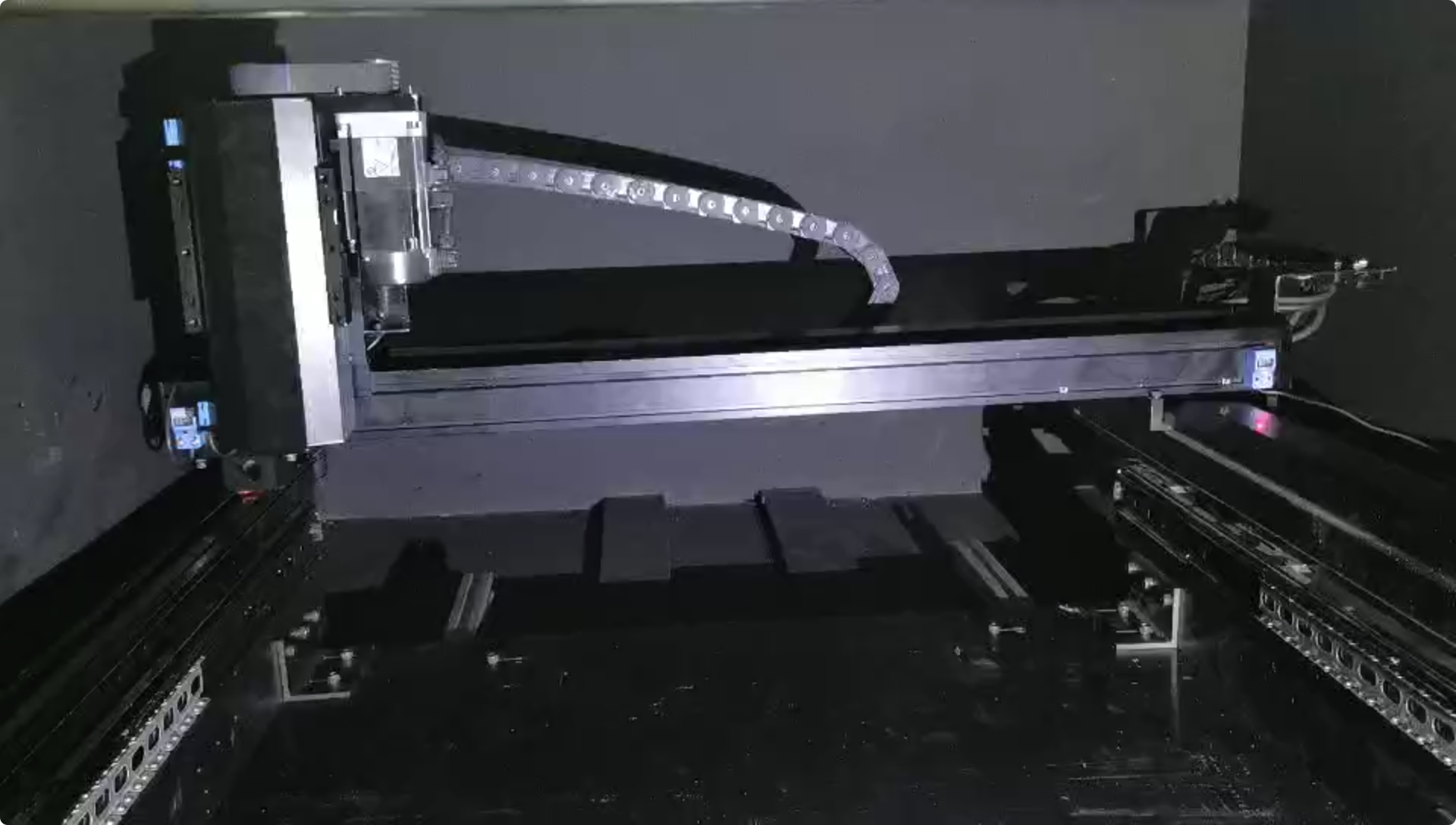}	
	\caption{\label{fig:j} A photograph of 3D servo motor.}
\end{figure}
As shown in figure \ref{fig:iii}, \ref{fig:i}, the batch test platform ~\cite{bt} includes a dark box (1220 mm $\times$ 1220 mm $\times$ 650 mm) and a PC. Inside the dark box, where the scintillator under test is placed, there are the servo motor, the electronic read-out system and a Sr90 source. The scintillator light yield is measured using the energy spectrum of Sr90 source.\par 
The 3D servo motor (figure \ref{fig:j}) are mounted in the dark box to carry the Sr90 radioactive source whose range of motion (800 mm $\times$ 800 mm $\times$ 300 mm) extends throughout the electronic board (530.5 mm $\times$ 530.5 mm). 

Inside the dark box, there is a sliding drawer motorized over the printed circuit board (PCB) which make the placement of scintillators much easier. The PCB consists of SiPMs and an electronic read-out system (4 SPIROC2E chips of 36 channels each, for a total of 144 channels), and the SiPM of each detector cell is directly welded on the board, and the placement of scintillators is based on the SiPM-on-Tile design (shown in Figure \ref{fig:ii}). \par
The SPIROC is a dedicated front-end electronics with SiPM readout, which realized in 0.35$\mu$m SiGe technology. It is an bi-gain, auto-triggered, 36-channel ASIC, with a range of 1 to 2000 p.e. by 12-bit Wilkinson ADC and the time with a 100ps accurate 12-bit TDC. An analog memory array with a depth of 16 for each is used to store the time information and the charge measurement. 
\begin{figure}[H]
	\centering
	\includegraphics[width=.5\textwidth]{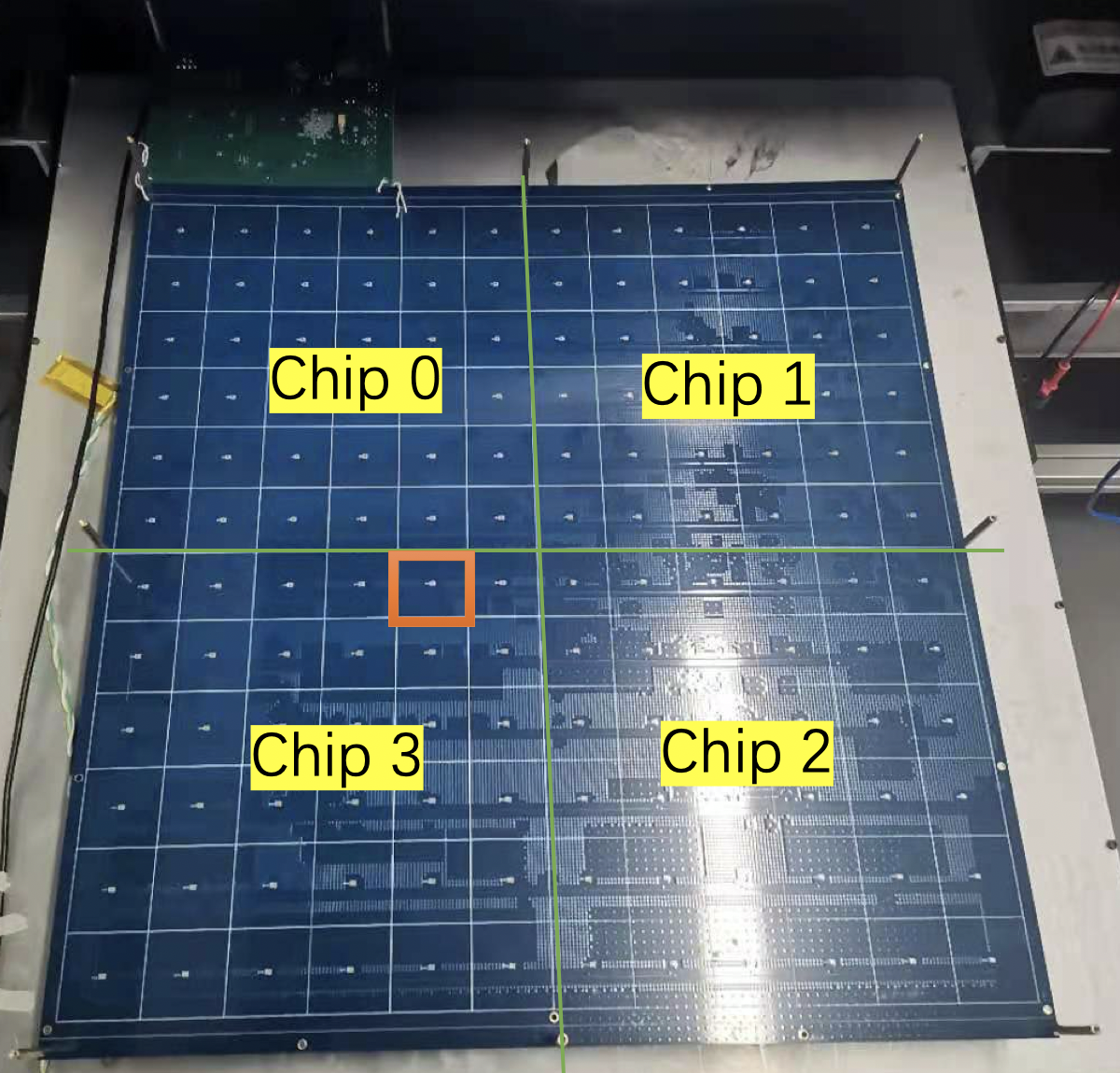}
	\includegraphics[width=.475\textwidth,angle=90]{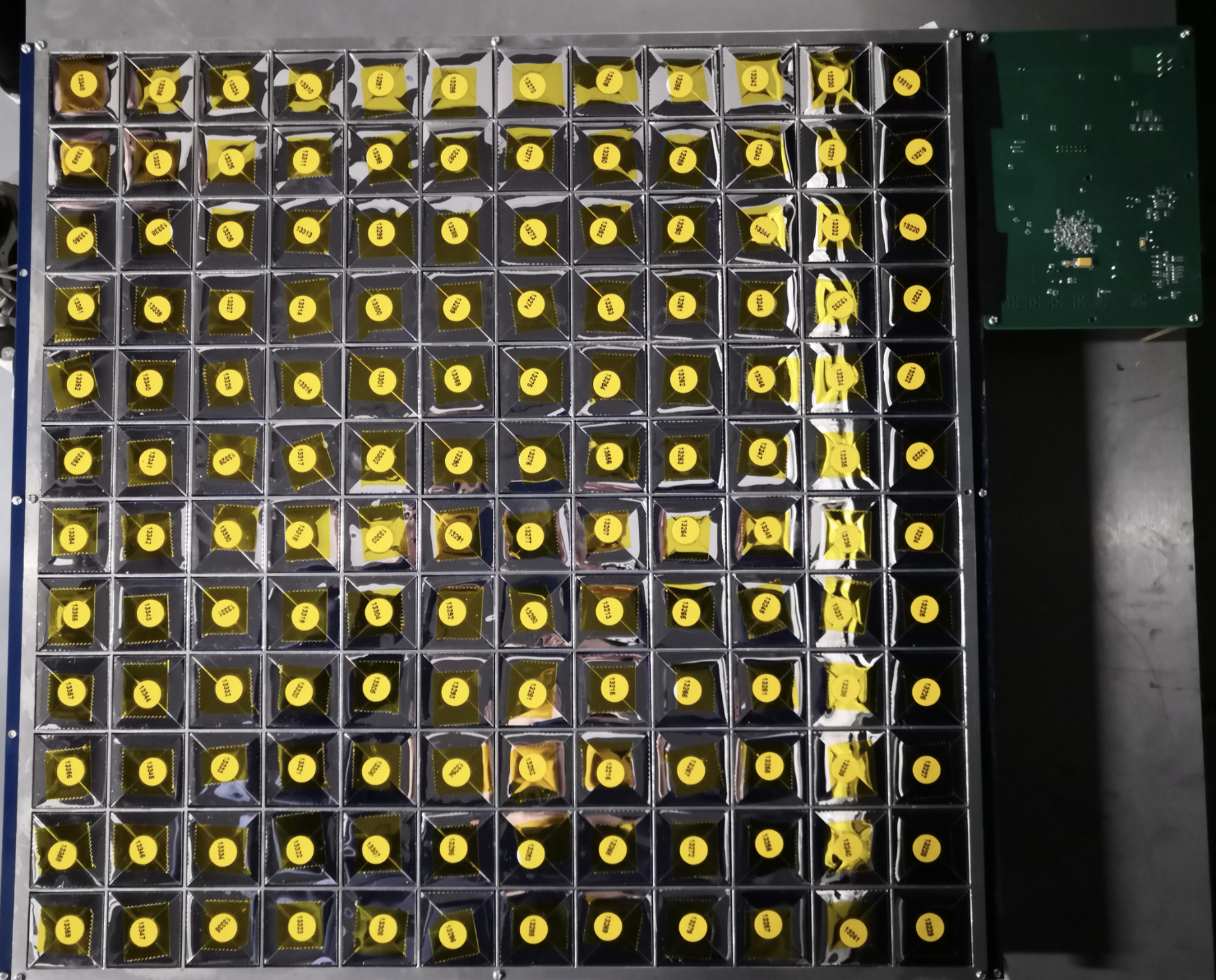}
	\caption{\label{fig:k}  A photograph of the PCB board. The 12 $\times$ 12 channels are read out by 4 SPIROC2E chips. The channel  inside the orange box is dead.}
\end{figure}
Figure \ref{fig:k} shows PCB board, the cross curve in green separate the PCB board into 4 part, each served by a SPIROC2E chip. Each small square area corresponds to one channel, with one SiPM (the small dot in figure) welded. The small green board is the DIF board with 6V power supply connected to PC by USB.

\subsection{Temperature and Humidity}
This test uses the HAMAMATSU SiPM S13360-1325PE, with dark count rate $7\times10^4$ to $21\times10^4$ at 0.5 p.e. threshold. The SiPM gain relies heavily on temperature ~\cite{m,n}. A SiPM contains a large number of pixels, each one consisting of a Geiger mode avalanche photodiode (APD) and a quenching resistance. The gain of SiPM is propositional to the reverse voltage (operating voltage) which is several volts higher than the breakdown voltage of APDs. The breakdown voltage of APDs increases with the rise of temperature, which will cause the decline of the gain ~\cite{n}. Raspberry Pi and humidity and temperature sensor (HTU21) were used to detect temperature and humidity. \par

\begin{figure}[H]
	\centering
	\includegraphics[width=.5\textwidth]{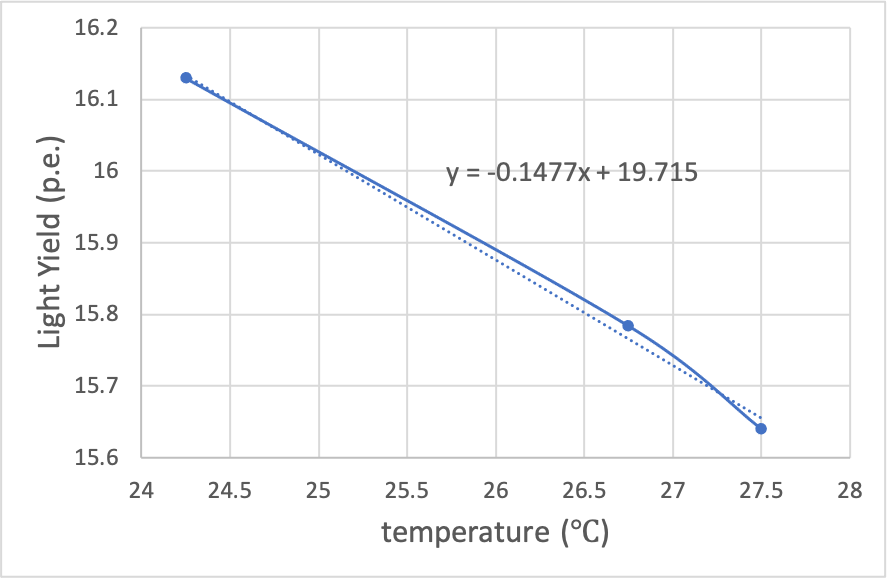}
	\caption{\label{fig:t} Measured light yield as a function of the temperature, the measurement is performed on chip 2 channel 18 with scintillator tile number 06939 taken as a reference.}
\end{figure}
From Figure \ref{fig:t}, take scintillator number 06939 as reference, we perform the measurement on chip 2 channel 18, the temperature recorded by Raspberry Pi and humidity and temperature sensor (HTU21), we got the measured light yield as a function of the temperature. The light yield is inversely proportional to the temperature (around -1\%/$^\circ$C, at 24 $^\circ$C, details see Figure \ref{fig:t}), and the influence of humidity can be ignored.

\subsection{Channel Calibration} 
\begin{figure}[h]
	\centering
	\includegraphics[width=.9\textwidth,page=3]{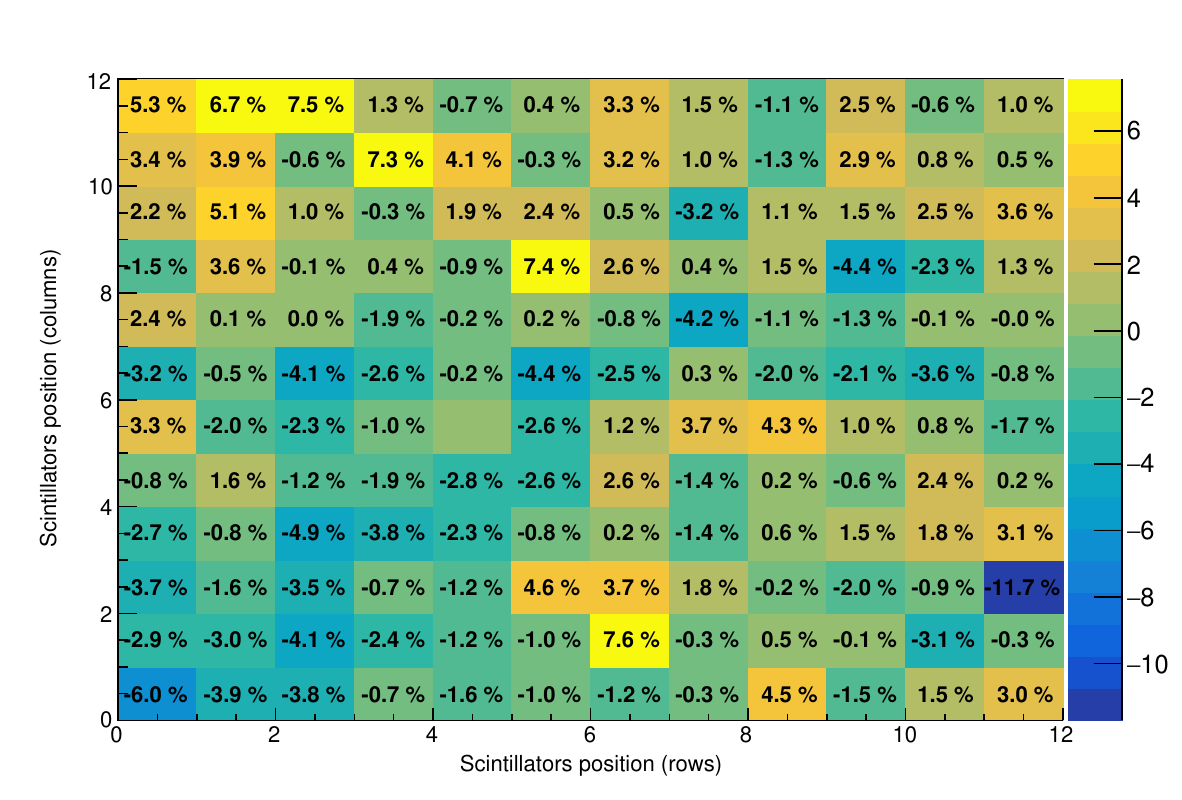}
	\caption{\label{fig:c} 144 channels calibration, the non-uniformity measurement. X-Y axis is the coordinate of scintillator, each bin corresponds to one channel in figure \ref{fig:k}, and the empty bin is the dead channel.}
\end{figure}

The scintillator light yield, SiPM gain, and SPIROC2E electronic gain are the 3 main sources of response non-uniformity. According to Section \ref{sec:intro}, in order to obtain reliable test results of scintillator's light yield, we calibrate the batch test system non-uniformity result from the SiPM gain non-uniformity, and the SPIROC2E electronic gain, result shown in figure \ref{fig:c}, which is the important calibration of the batch test system.

\section{Light yield measurement}
\begin{figure}[H]
	\centering
	\subfloat[]{\label{fig:baselinea}\includegraphics[width=.5\textwidth,page=238]{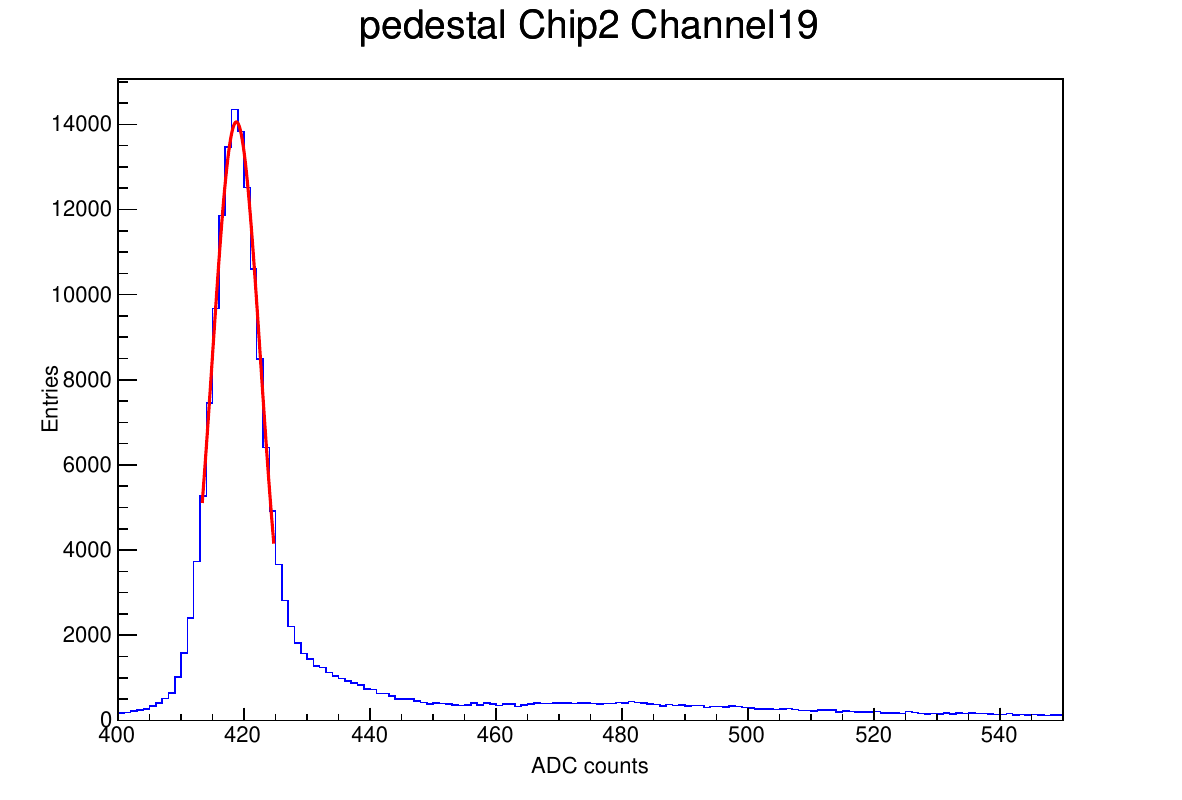}}
	\subfloat[]{\label{fig:baselineb}\includegraphics[width=.5\textwidth,page=297]{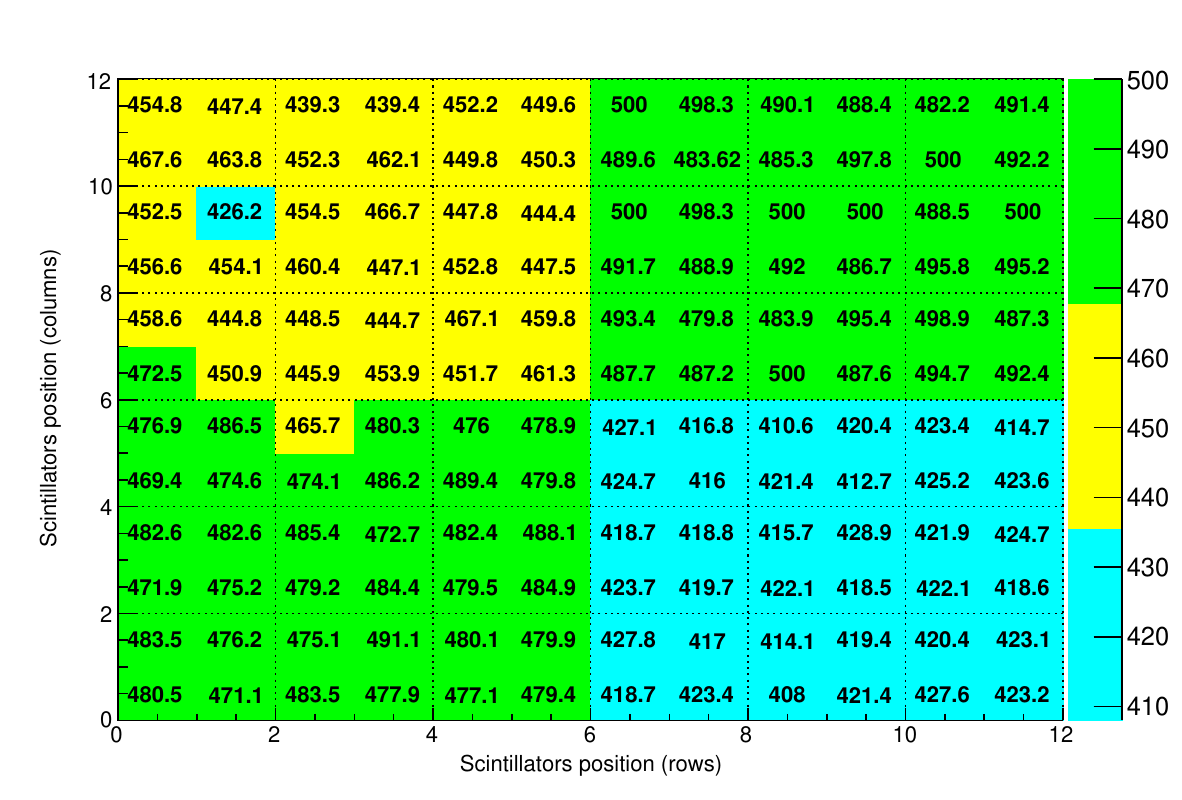}}
	\caption{\label{fig:baseline} (a) Pedestal of one channel. The blue line is the test data, the red line is Gaussian fit, and its mean (419.7 ADC) is taken as pedestal. (b) shows the pedestal distribution of 144 channels, 4 SPIROC2E which can be seen clearly except 3 individual channels.}
\end{figure}

\begin{figure}[H]
	\centering
	\subfloat[]{\label{fig:spga}\includegraphics[width=.5\textwidth,page=93]{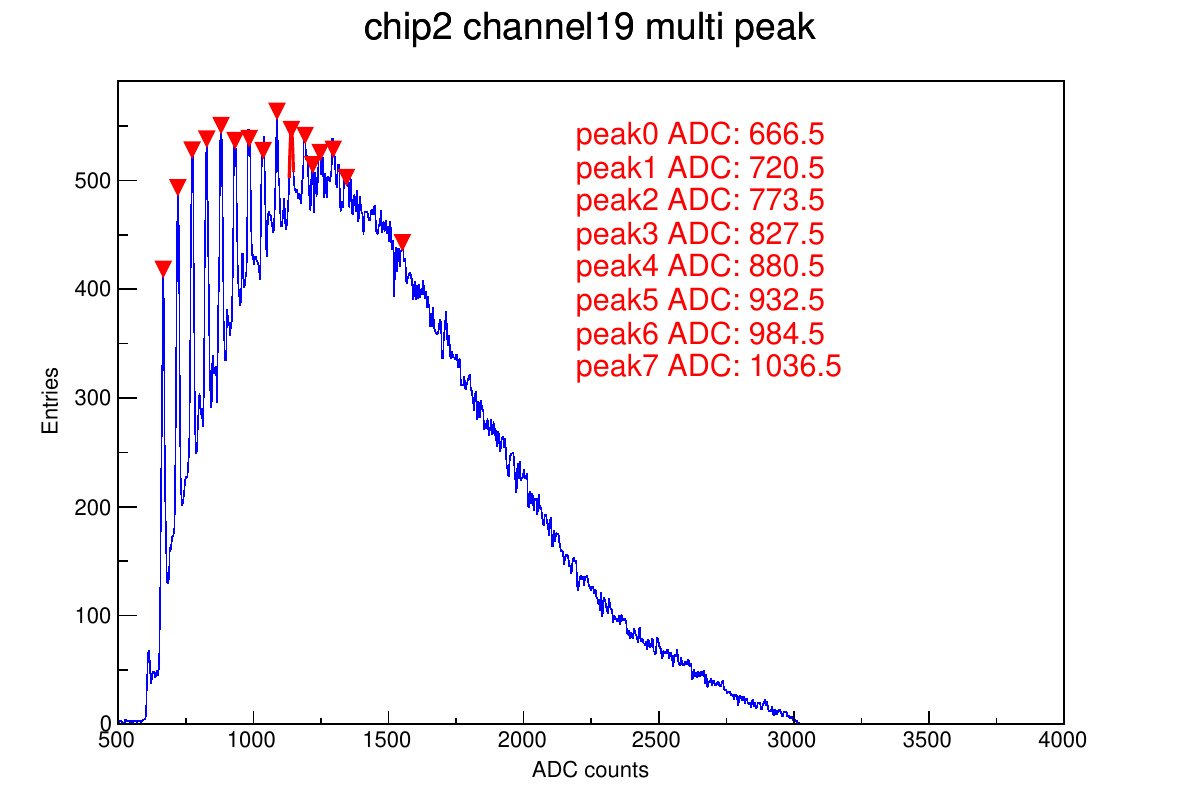}}
	\subfloat[]{\label{fig:spgb}\includegraphics[width=.5\textwidth,page=94]{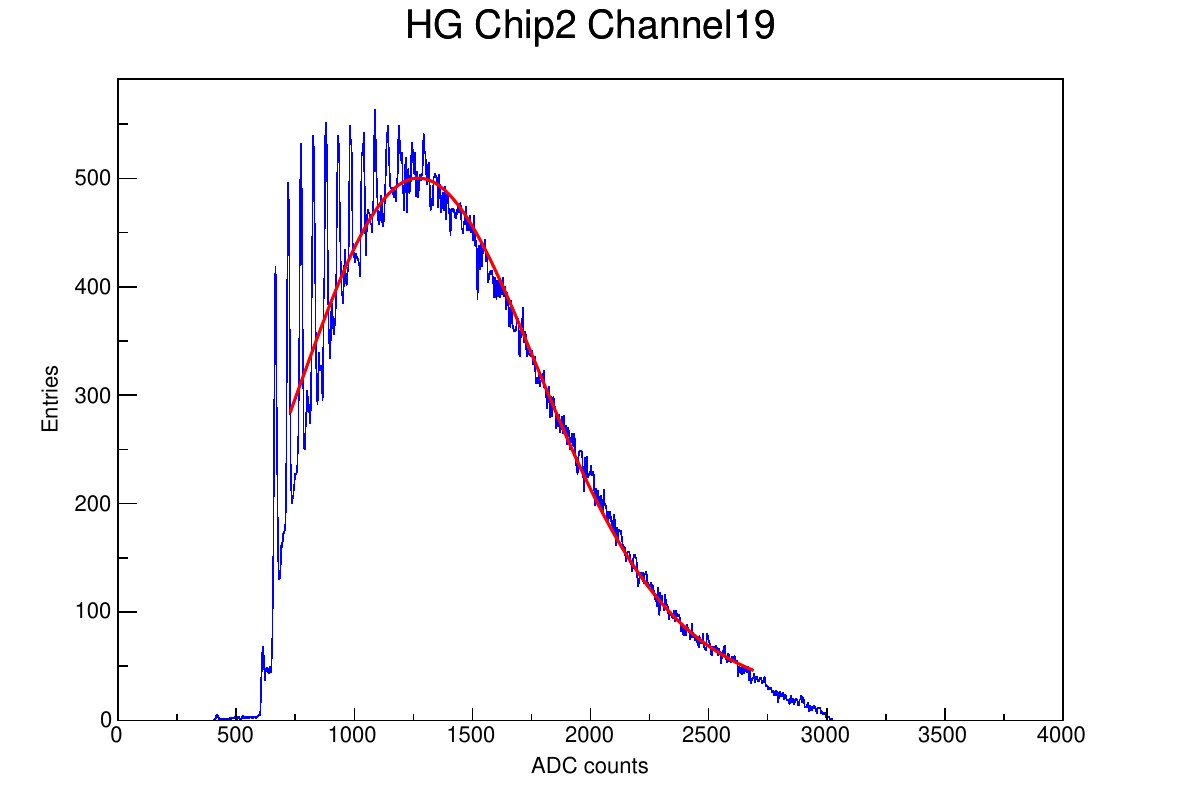}}
	\caption{\label{fig:spg} Single Photon Gain (a) and Peak of MIP (b) with Sr90.}
\end{figure}

\begin{figure}[H]
	\centering
	\includegraphics[width=.9\textwidth,page=18]{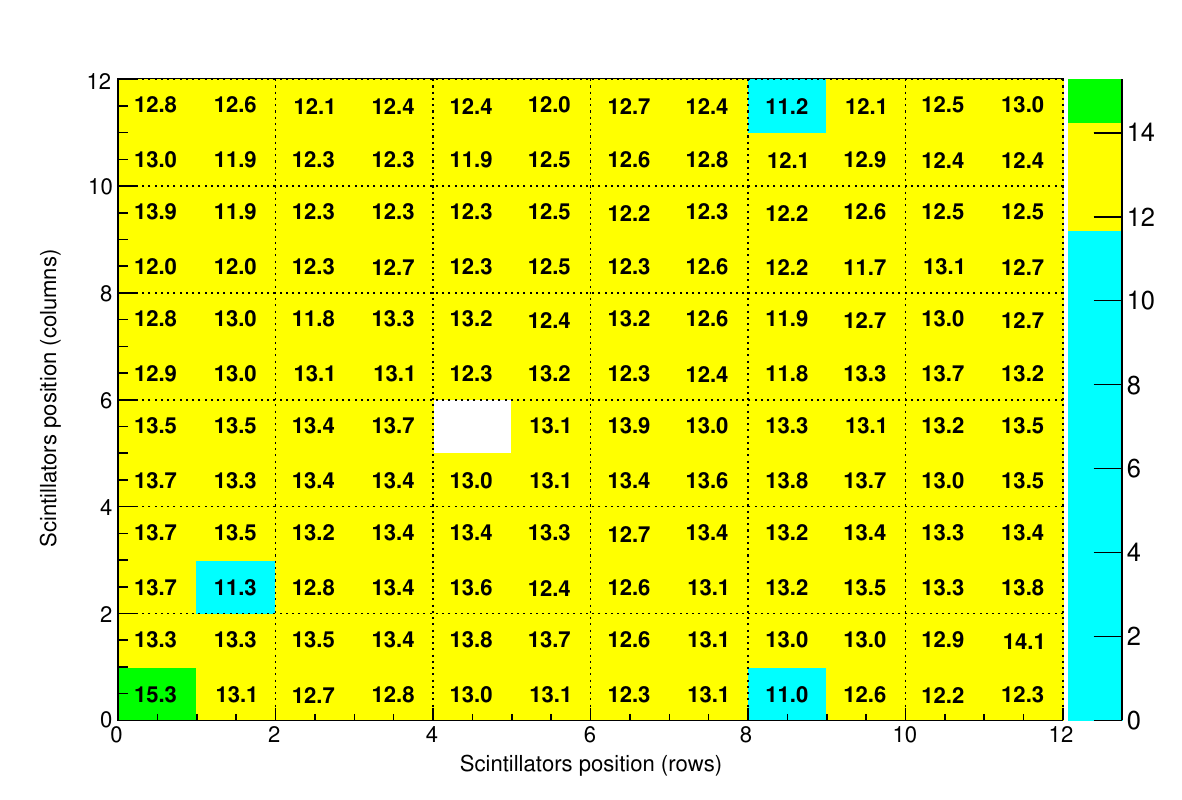}
	\caption{\label{fig:ly} The 2D map of scintillators' light yield (143 pieces p.e.). 140 pieces within 10\% (11.7 p.e. < LY < 14.3 p.e.). Two have a smaller light yield ( LY < 11.7 p.e.) and one has a larger light yield (LY > 14.3 p.e.). }
\end{figure}

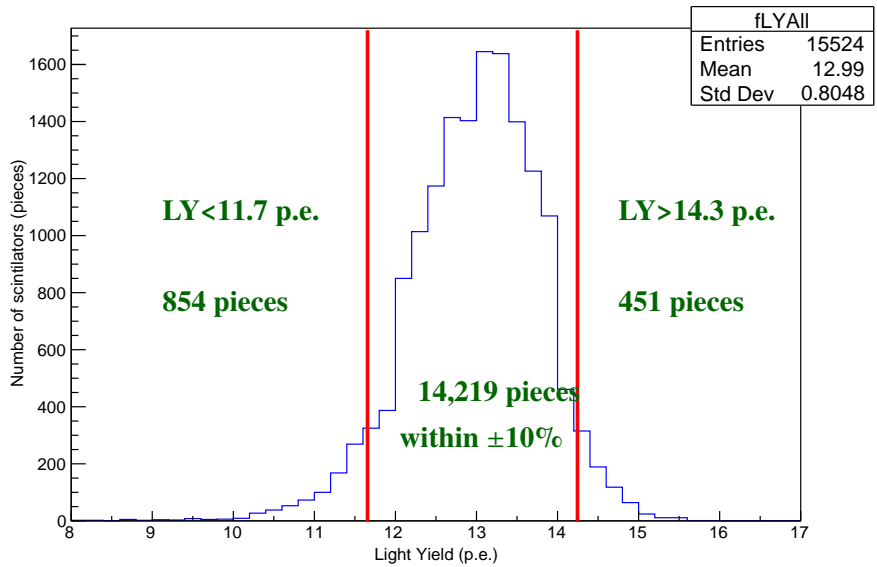
\begin{figure}[H]
	\centering
	\begin{overpic}[width=12cm,keepaspectratio,page=1,angle=0]{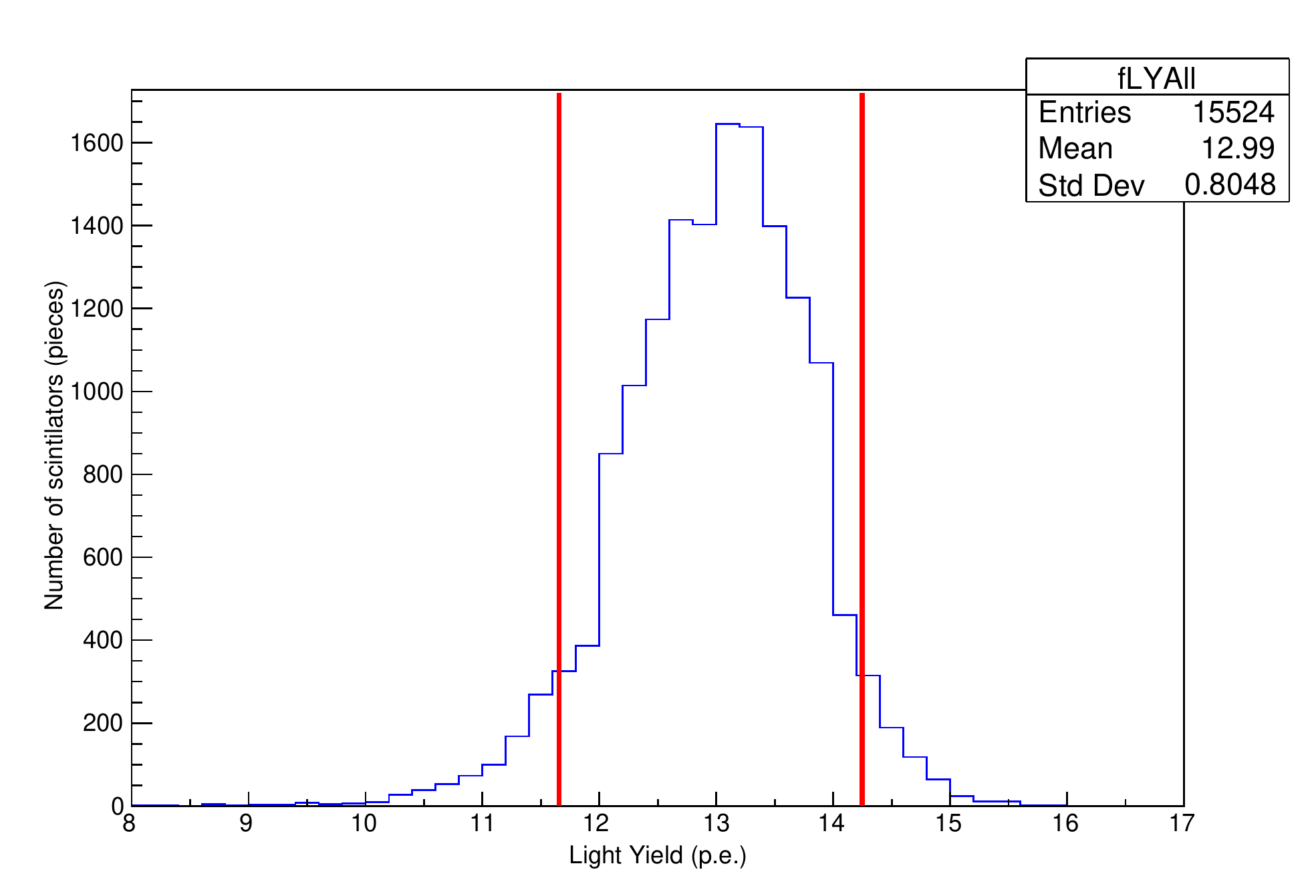}
		\put(20,40){\bfseries\color{DarkGreen}{LY<11.7 p.e.}}
		\put(20,30){\bfseries\color{DarkGreen}{854 pieces}}
		\put(70,40){\bfseries\color{DarkGreen}{LY>14.3 p.e.}}
		\put(70,30){\bfseries\color{DarkGreen}{451 pieces}}
		\put(48,20){\bfseries\color{DarkGreen}{14,219 pieces}}
		\put(46,15){\bfseries\color{DarkGreen}{within $\pm$10\%}}
	\end{overpic}
	\caption{\label{fig:u}Light yield distribution after calibration of 15524 scintillators. The mean value is 12.99 p.e., $\pm$10\% light yield window is at 11.7 p.e. < LY < 14.3 p.e (the red line along Y axis is the threshold line of 11.7 p.e./14.3 p.e. )}
\end{figure}

The light yield of each scintillator is determined by the peak of MIP, the single photon gain and the baseline, it's determined: 
\begin{equation}\label{eq:a} 
	LY=\frac{MIP-Baseline}{SinglePhotonGain} 
\end{equation}
Based on equation \ref{eq:a}, the peak of MIP test, the pedestal test and single photon gain detection are carried out separately. The pedestal is not a discrete value but a continuous value which satisfies Gaussian distribution. The mean of Gaussian fit is taken as the pedestal (see figure \ref{fig:baselinea}). Figure \ref{fig:baselineb} shows the distribution of 144 channels.
Single photon gain is the key factor  for calculating the light yield of tested scintillators. The SiPMs which are selected are HAMAMATSU S13360-1325PE, with outstanding photon counting capability, which make the peaks observable by setting an appropriate threshold for each channel. Figure \ref{fig:spga} shows more than 10 visible photon peaks, selected to find the single photon gain. Considering the Landau distribution of energy loss, the peak of MIP is fitted by Landau and Gauss convolution, result shown in Figure \ref{fig:spgb}. The chip 3 channel 4 with a good pedestal readout, demonstrate it's a SiPM dead channel.
Calculated by equation \ref{eq:a}, figure \ref{fig:ly} shows the light yield of 143 pieces in one batch. The light yield of scintillator in yellow bins have an uniformity of 10\% (11.7 p.e. < LY < 14.3 p.e. see figure \ref{fig:u}).


Figure \ref{fig:u} shows the result of 15524 pieces scintillators' light yield distribution after calibration with a non-uniformity of 0.8 (statistics standard deviation). There are 14219 pieces within $\pm$10\% light yield window. 

\section{Conclusions}
15524 scintillators in total were tested using a SPIROC2E-SiPM scintillator batch test platform for the AHCAL prototype of CEPC. The batch test platform includes a PC and a dark box, and inside the dark box, there are the servo motor, the radioactive source, the electronic readout system and temperature humidity monitors. The scintillator light yield is measured using the energy spectrum of Sr90 source. Single photon calibration, pedestal test and scintillator light yield measurements are carried out successfully in the platform. The channel non-uniform of this platform is calibrated and used to correct the light yield. Test results of this paper will be used to advise the scintillator selection of AHCAL prototype (13k channels needed). About 91.6\% of scintillators (14219 pieces) are qualified within 10\% of light yield window.

\acknowledgments
The study was supported by National Natural Science Foundation of China (No. 11961141006), National Key Programmes for S\&T Research and Development (Grant No.: 2018YFA0404300 and 2018YFA0404303).



\begin{thebibliography}{99}
\bibitem{a}
ATLAS Collaboration, \emph{Observation of a new particle in the search for the Standard Model Higgs boson with the ATLAS detector at the LHC}, In: Phys. Lett, B716 (2012).
arxiv: 1207.7214 

\bibitem{b}
CMS Collaboration, \emph{Observation of a new boson at a mass of 125 GeV with the CMS experiment at the LHC}, In: Phys. Lett. B716 (2012)
arxiv:1207.7235.

\bibitem{c}
Lyndon Evans and Philip Bryant, \emph{LHC Machine}, In: JINST 3 (2008), S08001. 

\bibitem{d}
Peter W. Higgs, \emph{Broken Symmetries and the Masses of Gauge Bosons}, In: Phys. Rev. Lett. 13 (1964)

\bibitem{e}
GS Guralnik, CR Hagen, and TWB Kibble, \emph{Global Conservation Laws and massless Particles},In: Phys. Rev. Lett. 13 (1964)

\bibitem{f}
Sheldon L. Glashow, \emph{Partial-symmetries of weak interactions}, In: Nuclear Physics 22.4 (1961)

\bibitem{g}
 CEPC ‐ SPPC Study Group. \emph{CEPC Conceptual Design Report Volume I-
Accelerator[R]}, 2018: 21, 25. 

\bibitem{h}
 CEPC ‐ SPPC Study Group. \emph{CEPC-SPPC Progress Report (2015 – 2016)
Accelerator[R]}, 2017: 1 

\bibitem{i}
CEPC Study Group, \emph{CEPC Conceptual Design Report: Volume 2-Physics \& Detector}
arXiv :1811.10545

\bibitem{j}
M.A. Thomson, \emph{Particle Flow Calorimetry and the PandoraPFA Algorithm}, Nucl. Instr. Meth. Phys. Res. A 611 (2009)
arXiv:0907.3577

\bibitem{k}
Marshall, J.S. et al, \emph{The Pandora Particle Flow Algorithm}
arXiv:1308.4537 (2013)

\bibitem{sci}
L. Li et al, \emph{Optimization of the CEPC-AHCAL scintillator detector cells}, Journal of Instrumentation, Volume 16, March 2021

\bibitem{sipm}
Jiechen Jiang et al, \emph{Study of SiPM for CEPC-AHCAL},  Journal of Instrumentation, Volume 980, August 2020

\bibitem{SP14} M Bouchel et al, \emph{SPIROC (SiPM Integrated Read-Out Chip): Dedicated very front-end electronics for an ILC prototype hadronic calorimeter with SiPM read-out}, Journal of Instrumentation, Volume 6, January 2011

\bibitem{l}
Marteinsdóttir, María., \emph{Light Yield Non-uniformity in PWO Scintillators}, Master’s thesis, Stockholm University (2009).

\bibitem{bt}
H. Liu et al, \emph{Development of a SPIROC2E-based scintillator test platform for CEPC AHCAL prototype}, Journal of Instrumentation, Volume 15, October 2020

\bibitem{m}
Ramilli, Marco., \emph{Characterization of SiPM: temperature dependencies}, 2008 IEEE Nuclear Science Symposium Conference Record. IEEE, (2008).


\bibitem{n}
Hamamatsu, \emph{How does temperature affect the gain of an SiPM?} \url{https://hub.hamamatsu.com/jp/en/technical-note/sipm-temperature-gain/index.html}



\end{thebibliography}
\end{document}